\documentclass{INTERSPEECH2023}
\usepackage{amsmath,times,booktabs,tabularx,amsmath,graphicx,url,cite,balance,amssymb, amsthm, array, multirow, dblfloatfix, float}

\interspeechcameraready

\title{A Multi-Task Learning Framework for Sound Event Detection using\\ High-level Acoustic Characteristics of Sounds}
\name{Tanmay Khandelwal and Rohan Kumar Das}

\address{
  Fortemedia Singapore, Singapore}
\email{f20170106p@alumni.bits-pilani.ac.in, rohankd@fortemedia.com}

\begin{document}

\maketitle
 
\begin{abstract}
 Sound event detection (SED) entails identifying the type of sound and estimating its temporal boundaries from acoustic signals. These events are uniquely characterized by their spatio-temporal features, which are determined by the way they are produced. In this study, we leverage some distinctive high-level acoustic characteristics of various sound events to assist the SED model training, without requiring additional labeled data. Specifically, we use the DCASE Task 4 2022 dataset and categorize the 10 classes into four subcategories based on their high-level acoustic characteristics. We then introduce a novel multi-task learning framework that jointly trains the SED and high-level acoustic characteristics classification tasks, using shared layers and weighted loss. Our method significantly improves the performance of the SED system, achieving a 36.3\% improvement in terms of the polyphonic sound event detection score compared to the baseline on the DCASE 2022 Task 4 validation set.
\end{abstract}

\noindent\textbf{Index Terms}: sound event detection, multi-task learning, high-level acoustic characteristics
\vspace{-2mm}

\section{Introduction}
\label{sec:intro}

Humans have the ability to perceive the environment by detecting and segregating various sound events in the surroundings. This can be a challenging task for machines~\cite{challengebook} and therefore, sound event detection (SED) research aims to replicate the human auditory system for robust automatic detection of sound events. The SED systems enable machines to recognize the onset and offset of sound events and classify them into different environmental contexts. In addition, SED can provide feedback to other systems, such as triggering an alarm. Thus, there has been significant research dedicated to developing novel methods for comprehending the sounds of everyday life in the context of practical applications such as smart homes~\cite{smarthome, smarthome2}, and audio surveillance~\cite{audiosurveillance2, babycrywork}.

For the SED models to accurately predict the temporal onset and offset, strongly labeled clips are a requirement to train the system in a supervised manner. The term “strongly labeled” refers to an audio clip in which events are annotated with their respective start and end times. Typically, audio segments outside the annotated start and end time limits are deemed non-target events and are not used as training data. The construction of this set of labeled data is associated with higher annotation costs and is also dependent on the annotator's subjective judgment. Thus, recent advancements have centered on reducing reliance on strongly labeled sets while achieving an effective system. These vary from simple data augmentation methods to semi-supervised learning methods~\cite{meanteacher, uda}. 

Data augmentation techniques, such as filter augmentation~\cite{fdy} and SpecAugment~\cite{specaug}, have been successful in improving the generalization capability of SED models. Filter augmentation weights frequency bands to extract useful information from a wider frequency range, while SpecAugment masks blocks of consecutive frequency channels and time frames in the spectrogram. In addition, a number of studies have demonstrated the efficacy of semi-supervised methods such as interpolation consistency training~\cite{skunit}, shift consistency training~\cite{pseudo}, and extensive self-training for generating pseudo-labels~\cite{stage, pseudoweak, noisy} to make use of weakly labeled as well as unlabeled sets and thereby reduce the need for strongly labeled sets.

Humans possess another ability to learn multiple tasks simultaneously and transfer knowledge gained from one task to another. Taking inspiration from this innate capability, multi-task learning (MTL) aims to jointly learn multiple related tasks, allowing the knowledge acquired in one task to benefit other tasks. In MTL, all tasks are treated equally, and the primary objective is to enhance the performance of each task. This approach promotes knowledge sharing among tasks, enhances the effectiveness of each task, and mitigates overfitting risk. As a result, MTL has emerged as a powerful tool for addressing the issue of data scarcity by leveraging existing knowledge and reducing the cost of manual labeling for learning tasks.

The characteristics of various sound occurrences that appear in nature are distinct, which can be used to aid the SED system in enhancing its generalization performance. For instance, sounds generated by mechanical or electrical devices tend to have a consistent mean and variance throughout the recording, unlike impulsive noises such as a dog barking. Previous studies, such as~\cite{postprocessing}, have used event duration information to tailor the length of the window size of the median filters to each sound event. Furthermore,~\cite{mtlscenes, asc1, asc2, asc3} have proposed analyzing methods of sound events and scenes jointly based on MTL for SED and acoustic scene classification. 

In this study, we propose a novel MTL framework that employs the high-level acoustic characteristics of sound events to group them under a few larger classes and use their classification as an additional task to support the primary SED system. The generated categories of classes serve as higher-level abstractions of the original sound classes, making them highly interrelated. To the best of our knowledge, this is the first study to incorporate high-level acoustic characteristics of sound events in an MTL framework, enabling the SED model to apply shared information acquired from an additional classification task. Furthermore, this MTL-based approach can be easily integrated into existing SED frameworks, and the additional branch for high-level acoustic characteristic classification (ACC) can be removed during inference, which will maintain the same number of model parameters during inference. We extend our work in~\cite{Khandelwal2022, ssp} by incorporating the proposed MTL method into the single SED branch to show its effectiveness. 

\section{Proposed MTL framework}
\label{sec:mtlframework}

\subsection{High-level acoustic characteristics}
\label{sec:motivation}
The world around us is filled with a vast array of sound events that vary in frequency content, duration, and profile. This rich source of acoustic information can be leveraged to train effective SED models. Using this simple idea to aid the existing SED models, we categorize the 10-classes in the DCASE 2022 Task 4 dataset into different categories based on the similarities in their high-level acoustic characteristics in this work. To use this additional information in the existing SED system, we further propose MTL-based joint learning. This allows the SED model to exploit the high-level acoustic characteristics of similar sound signals by adding another classification task to the single SED branch. The use of the MTL framework with joint training has the added benefit that once the MTL-based model is trained, the high-level ACC branch can be removed from the model architecture. As only the 10-class SED branch is used during the inference, the number of parameters remains the same as that of a single SED branch.  

\begin{figure}[t]
\centering  
\includegraphics[width=1\columnwidth]{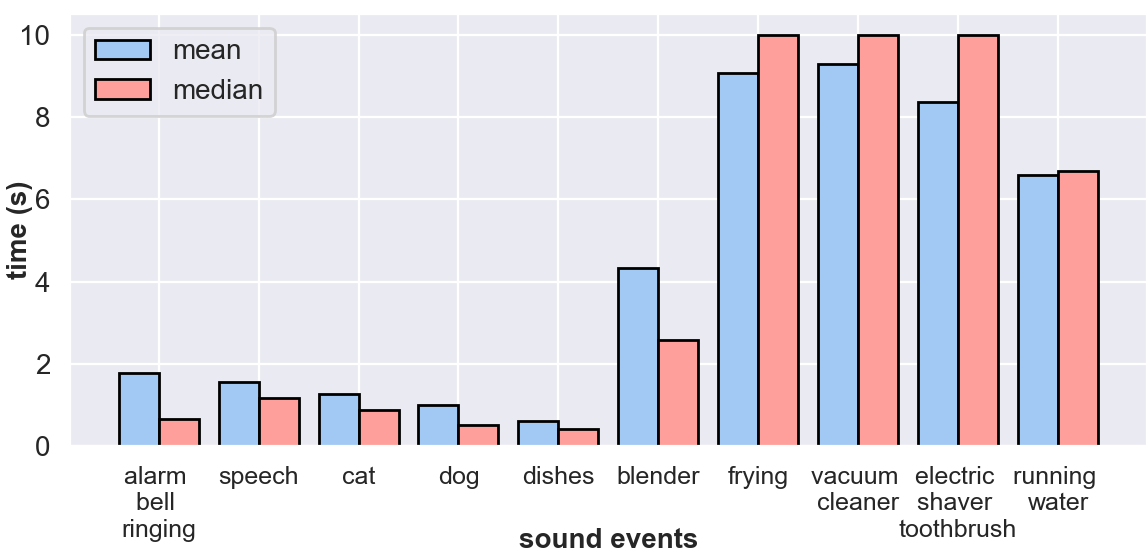}
\vspace{-6mm}
\caption{Bar plot showing mean and median duration for 10-classes in DCASE 2022 Task 4 strongly labeled training set.}
\label{fig:meanmed}
\vspace{-4mm}
\end{figure}

In this work, we concentrate on the 10-classes present in the DCASE 2022 Task 4 dataset. Through analysis of the strongly labeled set, we present the mean and median duration for each class in Figure~\ref{fig:meanmed}. Based on the average duration of the sound events, the plot indicates that these 10-classes can be grouped into two main categories. The 5-sound events on the right side of the bar plot have an average duration greater than 4s and can be referred to as “long duration events”, whereas those on the left have an average duration less than 2s and can be referred to as “short duration events”. Furthermore, we observe that the median duration for all “short duration events” is less than the mean, showing a positively skewed distribution, indicating that there are more short duration occurrences and fewer long duration occurrences. However, with the exception of “blender”, the median duration for the majority of “long duration events” is higher than the mean, showing a negatively skewed distribution, which suggests that there are more long duration occurrences and fewer short duration occurrences. 

\begin{figure}[t]
\centering  
\includegraphics[width=\columnwidth]{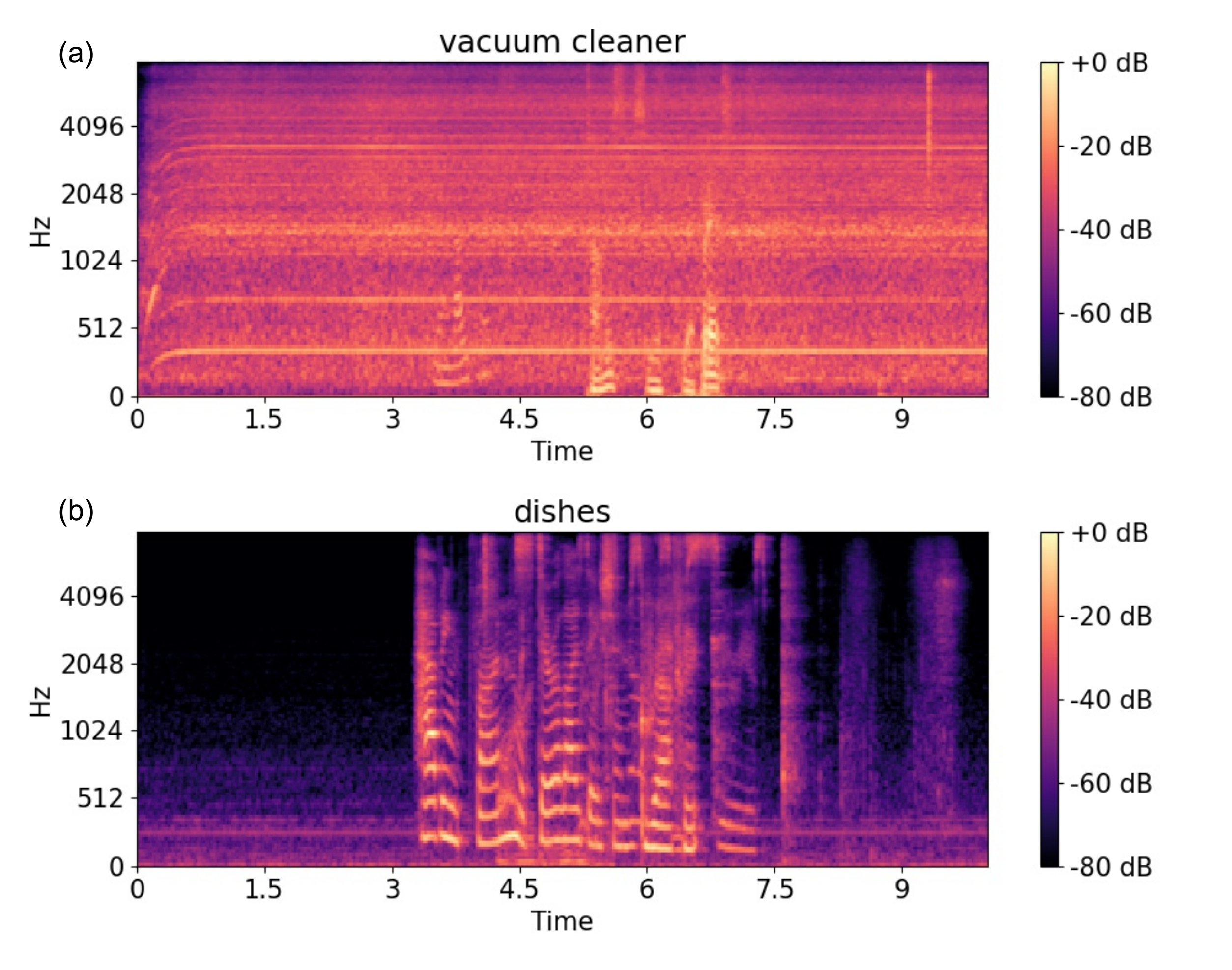}
\vspace{-8mm}
\caption{Log-mel spectrogram: (a) vacuum cleaner (b) dishes.}
\label{fig:logmel}
\vspace{-4mm}
\end{figure}

We further have a closer look at each of the two groups based on the duration statistics discussed above. On analyzing the “long duration events”, we observe from Figure~\ref{fig:logmel} (a) that the sound events such as “vacuum cleaner” are nearly stationary over time, exhibiting similar time-frequency trajectories. On the other hand, “short duration events” such as “dishes” fluctuate along the time axis, resulting in much variations along time-frequency trajectories. 

In general, the running motor of a “blender” and “vacuum cleaner” dominates over the sound of minor non-stationary noises such as the blender cutting a hard chunk or the vacuum cleaner's head hitting or rolling over other objects. Similarly, the sound of frying is classified as random noise because evaporation happens randomly and continuously, just as raindrops land randomly on the ground. As a result, all of these events have a relatively stable as well as less varying spectral content over time and can be approximated as stationary for a short period of time. Such events are classified as quasi-stationary~\cite{fdy, randomdata} and we categorize them into the acoustic characteristics-based class “A”. Similarly, sound events like “running water” and “electric shaver/toothbrush” can be considered to be quasi-stationary. However, in these instances, the sound produced by the “electric motor” and “running water” is not loud enough to drown out other impulsive sounds. These occurrences are therefore categorized as “semi-quasi-stationary”, as they share some characteristics with both stationary and non-stationary sounds, and we assigned them to class “B”.

\begin{figure}[t]
\centering  
\includegraphics[width=1\columnwidth]{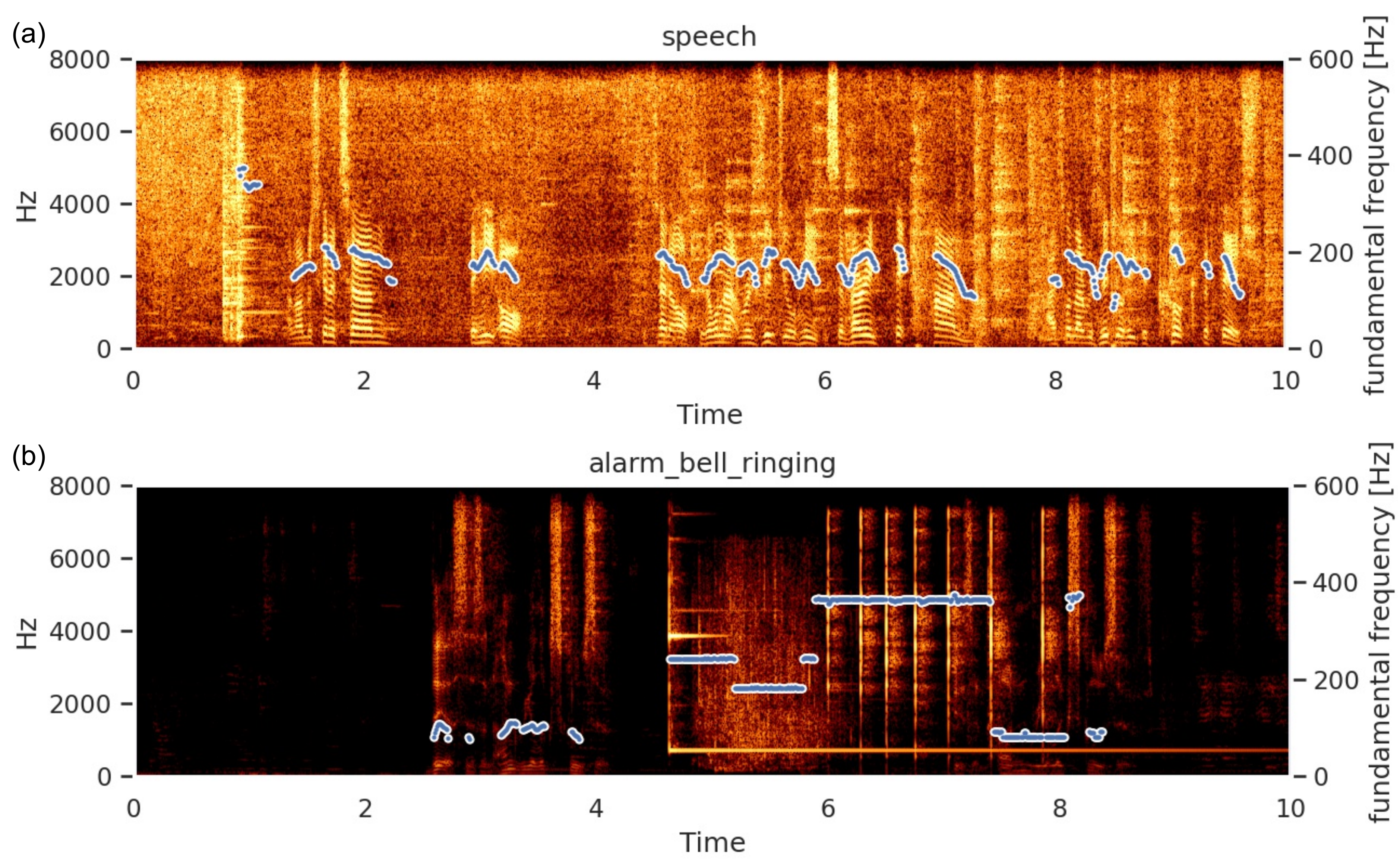}
\vspace{-6mm}
\caption{Pitch contours on a log-mel spectrogram: (a) speech (b) alarm bell ringing.}
\label{fig:pitchplot}
\vspace{-5mm}
\end{figure}

Along a similar direction, we investigate the “short duration events” and find that events like “dog”, “cat”, and “speech”, are all examples of vocalizations produced by living organisms. Thus, these events feature continuous changes in pitch due to the variations of their sound spectra, characterized by abrupt sounds like stops and momentary turbulent sounds like fricatives, as shown in the pitch contour plotted in Figure~\ref{fig:pitchplot} (a). Using this, we assign these events to class “C”. On the other hand, classes such as “dishes” and “alarm bell ringing” are characterized by brief and sudden sounds that exhibit a more consistent and stable pitch with less variability over time, as demonstrated in Figure~\ref{fig:pitchplot} (b). Additionally, compared to class “C”, the pitch of these sounds lies in a higher frequency region. Thus, both of these events are then assigned to class “D”. A summary of the sub-categorization of the 10 sound events of DCASE 2022 Task 4 based on the high-level acoustic characteristics discussed here is depicted in Table~\ref{table:categorization}.  

\begin{table}[t!]
\begin{center}
\caption{Categorization of the 10 sound events in DCASE Task 4 dataset into 4 classes based on the acoustic characteristics.}
\vspace{-2mm}
\label{table:categorization}
\resizebox{\columnwidth}{!}{%
\begin{tabular}{|l|c||l|c|}
\hline
  \textbf{Long duration events} &   \textbf{Class} &   \textbf{Short duration events} &   \textbf{Class}\\
\hline\hline
  Vacuum cleaner  &  A &  Speech &  C\\
\hline
  Frying  &  A &  Dog &  C\\
\hline
  Blender  &  A &  Cat &  C\\
\hline
  Electric shaver/toothbrush &  B &  Dishes &  D\\
\hline
  Running water  &  B &  Alarm bell ringing &  D\\
\hline
\end{tabular}}
\end{center}
\vspace{-8mm}
\end{table}

\subsection{Proposed method}
\label{sec:method}
In our proposed MTL framework, we use the high-level acoustic characteristics-based class categorization discussed in the previous section to add another task to the existing single SED framework. In particular, we assign the provided clip section to one of the high-level acoustic characteristics-based classes (“A”, “B”, “C” or “D”) and one of the 10 sound events. In order for the network to acquire low-level features that could enhance the performance of the SED model, these two tasks are built to share some common layers. These shared layers facilitate the extraction of features common to both SED and high-level ACC tasks. It has also been shown that when tasks vary in difficulty, the harder task's performance is improved by using the knowledge from the easier task~\cite{mtleasy}. Here, we consider SED to be the more difficult task and high-level ACC to be the simpler task, and thus we anticipate that the ACC will aid in the SED performance. Therefore, to conduct the joint training of these two tasks, we use a combined loss function $L_{MTL}$, which is the weighted loss function from the SED branch and the ACC branch. It can be expressed mathematically as
\vspace{-1mm}
\begin{equation}
\mathit{L_{MTL}} = \alpha \times L_{SED} + (1-\alpha) \times L_{ACC}
\vspace{-1mm}
\end{equation}

where $\alpha$ is the trade-off factor that regulates the weighted loss. $L_{SED}$ and $L_{ACC}$ correspond to the 10-class SED loss and the 4-class high-level ACC loss, respectively. Setting $\alpha$ = 1 is equivalent to a single SED branch. Furthermore, the ACC branch is removed during inference time. As a result, the number of parameters is the same as in the single SED branch.

\subsection{Proposed architecture}
\label{sec:architecture}
\begin{figure}[!t]
\centering  
\includegraphics[width=0.9\columnwidth]{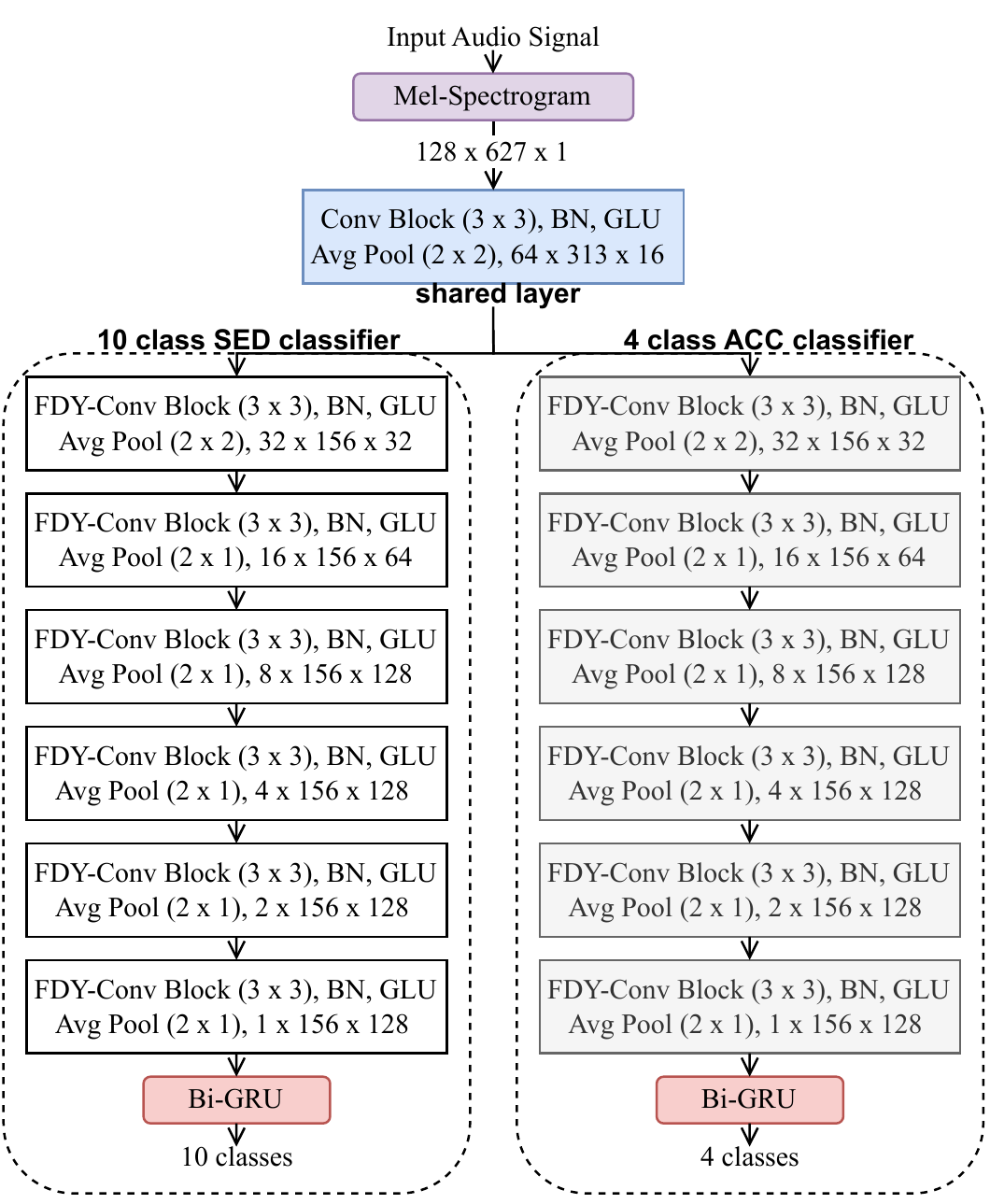}
\caption{Proposed MTL framework, with separate branches for 10-class SED task and 4-class high-level ACC task.}
\label{fig:mtlmodel}
\vspace{-1mm}
\end{figure}

We extend the work carried out by us in~\cite{Khandelwal2022, ssp} to transform it into the MTL framework mentioned in the previous subsection. Figure~\ref{fig:mtlmodel} illustrates its two branches, one of which corresponds to the 10-class SED task and the other to the 4-class high-level ACC task. Each of these branches employs the same frequency dynamic (FDY)-convolutional recurrent neural network (CRNN)~\cite{fdy} architecture on both sides after sharing a single convolutional block, which is referred to as “shared layer”, as shown in Figure~\ref{fig:mtlmodel}. 

\section{Experimental setup}
\label{sec:setup}

\subsection{Dataset and feature extraction}
\label{ssec:dataset}
We use 10-second audio clips from the DCASE 2022 Task 4 dataset taken from AudioSet~\cite{audioset} or synthesized with isolated sound events~\cite{fsdk50} and backgrounds~\cite{sins, tut} to model a domestic environment. The development set for this dataset is summarized in Table~\ref{tab:dataset}.

\begin{table}[t]
\begin{center}
\caption{Summary of DCASE 2022 Task 4 development set.}
\vspace{-2mm}
\label{tab:dataset}
\begin{tabular}{|l|l|l|}
\hline
\textbf{Set} & \textbf{Clips} & \textbf{Description} \\
\hline\hline
 \multirow{2}{*}{Strong} & 10,000  & Synthetic clips  \\\cline{2-3}
& 3,470  & Real recordings \\
\hline
Weak & 1,578 & Real recordings   \\
\hline
Unlabeled & 14,412 & Real recordings in domain \\
\hline 
\multirow{2}{*}{Validation} & 1,168 & Real strongly labeled clips \\\cline{2-3}
 & 2,500 & Synthetic strongly labeled clips \\
 \hline 
\end{tabular}
\end{center}
\vspace{-6mm}
\end{table}

We label each entire clip or event segment in the training set with the additional high-level acoustic characteristics-based classes (“A”, “B”, “C” or “D”). 
Then, all these audio clips are re-sampled at 16 kHz to a mono channel using Librosa. From the re-sampled clips, log-mel filter banks are extracted using a window size of 2048 samples with a hop length of 256 samples. In addition, to make feature sequences the same length, we zero-pad shorter sequences and truncate longer sequences. Then the mel-spectrogram of the clip is normalized to zero mean and unit variance is computed over the training set. 

\subsection{Two-stage system (TSS) for SED}
\label{ssec:twostage}
We used our two-stage method designed in~\cite{Khandelwal2022, ssp} for DCASE 2022 Task 4. This system has two stages: Stage-1 concentrates on audio-tagging (AT), and Stage-2 improves SED performance by using the reliable pseudo-labels generated by Stage-1. In Stage-1, the CNN-14-based pre-trained audio neural network~\cite{panns} is utilized as the feature extractor, whereas in Stage-2, FDY-convolutions are utilized. Both of these feature extractors are followed by 2 layers of bidirectional gated recurrent units~\cite{gru} with 1024 and 128 hidden units, respectively. The AT system (Stage-1) is used to make predictions on the unlabeled set and generate a pseudo-weak set. During the training of Stage-2, the weak and pseudo-weak sets are combined, and the strongly labeled and combined weakly (weak and pseudo-weak sets) labeled samples are given a weight of 1. In our study, we extend Stage-2 employing FDY-CRNN to construct the MTL framework described in Section~\ref{sec:architecture}, and assigning the high-level acoustic characteristics-based classes to each segment in the strongly labeled and combined weakly labeled set. 
 
\subsection{Training and post-processing}
\label{ssec:preprocessing}
With a similar training setting as in~\cite{Khandelwal2022, ssp}, we adopt a batch size of 48 for all experiments. Similarly, the model is trained using the Adam optimizer, with a learning rate ramp-up occurring over the first 50 optimization epochs and a maximal learning rate of 0.001. Stage-1 is trained using a total of 100 epochs, and Stage-2 is trained using a total of 200 epochs. Additionally, we also use the same data-augmentation techniques outlined in~\cite{Khandelwal2022, ssp} at each stage to enhance the model's generalizability. Since each event class has a different duration length, we also adopt the class-wise median filter~\cite{postprocessing} and search for the optimal filter length. The system was developed using PyTorch Lightning and trained using an NVIDIA Quadro RTX 5000 GPU.

\subsection{Evaluation metric}
\label{ssec:evaluation}
We used polyphonic sound event detection scores (PSDS)~\cite{psds} as a performance metric to evaluate the systems because it can surmount the limitations of conventional event F-scores based on collars. This is computed based on two distinct scenarios, with Scenario-1 denoted by PSDS1 concentrating on temporal localization and Scenario-2 denoted by PSDS2 focusing on class confusion prevention rather than reaction time. By using PSDS as our performance metric, we aim to provide a more nuanced evaluation of our SED models that accounts for both temporal localization and class confusion prevention.

\section{Results and analysis}
\label{sec:results}

\begin{table}[t]
\begin{center}
\caption{Performance comparison showing the importance of the proposed method on the DCASE 2022 Task 4 validation set.}
\vspace{-2mm}
\label{table:proposedcomp}
\resizebox{\columnwidth}{!}{%
\begin{tabular}{|c|c|c|c|}
\hline
\hfil \textbf{System} & \hfil \textbf{PSDS1} & \hfil \textbf{PSDS2}& \hfil \textbf{PSDS1+PSDS2}\\
\hline
\hline
\hfil Baseline: CRNN~\cite{baseline}  & \hfil0.351 & \hfil0.552 & \hfil0.903\\
\hline
\hfil Two-stage system (TSS)~\cite{Khandelwal2022, ssp} & \hfil0.472 & \hfil0.721 & \hfil1.193\\
\hline
\hline
\hfil TSS + MTL($\alpha$=0.5) & \hfil0.476 & \hfil0.751 & \hfil1.227\\ 
\hline
\hfil TSS + MTL($\alpha$=0.6) & \hfil0.457 & \hfil0.740 & \hfil1.197\\ 
\hline
\hfil TSS + MTL($\alpha$=0.7) & \hfil0.479 & \hfil0.738 & \hfil1.217\\ 
\hline
\hfil TSS + MTL($\alpha$=0.8) & \hfil0.480 & \hfil0.751 & \hfil \textbf{1.231}\\ 
\hline
\hfil TSS + MTL($\alpha$=0.9) & \hfil0.490 & \hfil0.729 & \hfil1.219\\ 
\hline
\hline
\hfil TSS + R-MTL($\alpha$=0.8) & \hfil0.461 & \hfil0.713 & \hfil 1.174\\ 
\hline
\end{tabular}
}
\vspace{-7mm}
\end{center}
\end{table}
This section presents the results and analysis of our various studies conducted in this work. Our experimental evaluation begins with an assessment of the SED system's effectiveness using the MTL framework. We use the two-stage system (TSS) defined in Section~\ref{ssec:twostage} and extend the single SED branch in Stage-2 to construct the proposed MTL framework. Since the primary task is the 10-class SED task, we assigned more weight to $L_{SED}$ by varying the trade-off factor, $\alpha$, from 0.5 to 0.9. The PSDS1 and PSDS2 obtained for each setup are shown in Table~\ref{table:proposedcomp}. It can be observed that the baseline adopted from DCASE 2022 Task 4~\cite{baseline} and the TSS~\cite{Khandelwal2022, ssp} developed by us achieves a total PSDS (PSDS1 + PSDS2) of 0.903 and 1.193, respectively. We find an improvement in the performance of the TSS by incorporating our proposed MTL framework, with the best performance of total PSDS of 1.231 when $\alpha$ = 0.8. Furthermore, with this method of joint training, the total parameters during the time of inference remain the same at 2.8M, as for the single SED branch. 

To test whether the proposed method of classifying classes based on high-level acoustic characteristics works, we run another experiment in which we classify highly different acoustic events into the same high-level classes and term it randomized-MTL (R-MTL). Table~\ref{table:randomcategorization} summarizes our categorization to retain similar acoustic classes into different high-level classes. According to the findings in Table~\ref{table:proposedcomp}, the total PSDS for TSS+R-MTL at the same $\alpha=0.8$ value as the best-performing system decreases significantly from 1.231 to 1.174. As a result, we can conclude that the method of categorizing sound events has a substantial impact on performance, and our method described in Section~\ref{sec:motivation} aids in the SED performance using MTL.

We then demonstrate the effectiveness of our proposed approach on the development set by comparing it to the top teams from DCASE 2022 Task 4 with single systems (no ensembling). The reported performances of other systems are taken directly from their respective technical reports. Table~\ref{table:single} shows the comparison of our proposed system to those top-performing team's single systems in a sorted order based on total PSDS value. We observe that our proposed MTL framework outperforms most of the systems and comes really close to the best single system with a larger model size of 15.4M parameters and multiple stages of self-training. It is also worth noting that our proposed framework can be easily integrated into other SED systems with just an additional branch for the high-level ACC task, without increasing the number of parameters during inference.

\begin{table}[t!]
\begin{center}
\caption{Categorization of the 10 sound events in DCASE Task 4 dataset into 4 classes to keep similar events in separate high-level acoustic characteristics classes.}
\vspace{-2mm}
\label{table:randomcategorization}
\resizebox{\columnwidth}{!}{%
\begin{tabular}{|l|c||l|c|}
\hline
  \textbf{Sound events} &   \textbf{Class} &   \textbf{Sound events} &   \textbf{Class}\\
\hline\hline
 Alarm bell ringing &  A &  Dishes &  C\\
 \hline
 Blender &  A &  Frying &  C\\
 \hline
 Electric shaver/toothbrush &  A &  Speech &  C\\
 \hline
 Vacuum cleaner &  B &  Running water &  D\\
 \hline
 Dog & B & Cat &  D\\
 \hline
\end{tabular}}
\end{center}
\vspace{-4mm}
\end{table}

\begin{table}[t]
\begin{center}
\caption{Comparison with top-ranked single systems (without ensemble) from DCASE Task 4 2022 on the validation set.}
\vspace{-2mm}
\label{table:single}
\setlength\extrarowheight{3pt}
\resizebox{\columnwidth}{!}{%
\begin{tabular}{|l|c|c|c|c|}
\hline
  \textbf{System} &   \textbf{PSDS1} &   \textbf{PSDS2} &   \textbf{PSDS1+PSDS2} & \textbf{\#Parameters}\\
\hline\hline
  Ebbers-UPB-task4~\cite{ebbers}  &  0.505 &  0.807 & 1.312 & 15.4M\\
\hline
 \textbf{TSS + MTL} ($\alpha$=0.8)  &  \textbf{0.483} &  \textbf{0.728} & \textbf{1.231} &  \textbf{2.8M}\\
\hline
  Zhang-UCAS-task4~\cite{ucas}  & 0.459 &  0.672 & 1.131 &  11M\\
\hline
  Kim-GIST-task4~\cite{kim}   &  0.455 &  0.670 & 1.125 & 1M\\
\hline
  Dinkel-XiaoRice-task4~\cite{dinkel}  &  0.425 &  0.644 & 1.069 &  37M\\
\hline
\end{tabular}}
\vspace{-8mm}
\end{center}
\end{table}

\section{Conclusion}
\label{sec:conclusion}
In this paper, we propose a novel MTL framework that leverages high-level acoustic characteristics to enhance the SED performance, without usage of additional annotated data. Our proposed framework performs joint learning with an additional high-level ACC task branch using a weighted loss, with both tasks sharing common layers to leverage cross-task information by sharing parameters (weights). The studies revealed that the MTL framework with additional high-level ACC task improves the primary SED system's performance. We also demonstrate the effectiveness of our proposed sound event categorization by separating similar sound events into distinct high-level acoustic characteristics classes and comparing our system's performance with the top teams in DCASE 2022 Task 4 on the validation set. In the future, we intend to incorporate an adaptive trade-off factor to control the weighted loss. 

\newpage
\balance
\bibliographystyle{IEEEtran}
\bibliography{mybib}
\end{document}